\documentclass{iopart}
\usepackage{iopams,epsfig}

\begin{document}
\title
{New exactly solvable periodic potentials for the Dirac equation}

\author{
B F Samsonov\dag %
\footnote[2]{To whom correspondence should be addressed
(boris@metodos.fam.cie.uva.es)} \footnote[3]{On leave from Tomsk
State University, 634050 Tomsk, Russia.
 },\
A~A~Pecheritsin$\|$,\ E~O~Pozdeeva$\|$ \
 and  M~L~Glasser$\P$ }

\address{\dag\ Departamento de F\'{\i}sica Te\'orica, Universidad de
Valladolid,  47005 Valladolid, Spain}

\address{$\|$ \ Department of Quantum Field Theory,
Tomsk State University, Lenin Ave, 634050 Tomsk, Russia}

\address{$\P$\ Department of Physics and Center for Quantum Device Technology,
Clarkson University, Potsdam NY 13699-5820 (USA)}

\ead{\mailto{samsonov@phys.tsu.ru}, \mailto{pecher@ido.tsu.ru},
\mailto{laryg@clarkson.edu}
}

\begin{abstract}
A new exactly solvable relativistic periodic potential is obtained
by the periodic extension of a well-known transparent scalar
potential. It is found that the energy band edges are determined
by a transcendental equation which is very similar to the
corresponding equation for the Dirac Kronig-Penney model. The
solutions of the Dirac equation are expressed in terms of
elementary functions.
\end{abstract}

\pacs{1315, 9440T}

\vspace{4em}

\address{European Journal of Physics 24 (2003) 435-441}

\newpage

\section{Introduction}

The properties of a relativistic particle in a one-dimensional
periodic structure play an important role in understanding many
phenomena in Solid State and Nuclear
Physics
 \cite{Mendez-1991,McKellar-1987a}.
For this purpose the relativistic generalization of the classic
Kronig-Penney model
 \cite{McKellar-1987a,McKellar-1987b}, the
 numerical solution of the Dirac equation
 \cite{Mendez-1991}
and a class of exactly solvable nonlocal separable periodic
potentials \cite{Glasser-1983} have been studied. Nevertheless,
due to its elementary character, the Dirac Kronig-Penney model
remains the most frequently used model for studying
one-dimensional relativistic band structures.

A new method for constructing exactly solvable non-relativistic
periodic potentials has been recently proposed
\cite{ejp_2001_v22_p22}.
Starting from an easily solved
(non-periodic) potential, a new solvable potential is generated by
using the familiar
 Darboux transformation.
 This new potential is then considered on a bounded interval and its periodic extension is
constructed. In this paper we generalize this procedure to
construct a solvable scalar potential for the Dirac equation.

\section{Band structure for a periodic potential}

Here we recall the main features of the Dirac equation with a
periodic potential (see e.g.
\cite{McKellar-1987a,McKellar-1987b}).

Consider the one-dimensional Dirac equation
\begin{equation}
h \psi(x) = E \psi(x), \quad h = i \sigma_y \partial_x + V(x)\,,
\label{dirac}
\end{equation}
where $V(x)$ is a real self--adjoint matrix potential, which we
assume has the canonical representation (for more details see e.g.
\cite{Levitan})
\begin{equation}
V(x) = p(x) \sigma_z + q(x) \sigma_x\,.
\end{equation}
$\sigma_{x,y,z}$ are the usual Pauli matrices and
$\psi(x)=(\psi_1(x),\psi_2(x))^t$
 (the superscript ``$t$" meaning the transpose)
 is a two-component spinor wave function. For any
real value of the parameter $E$ equation \eref{dirac} has two real
linearly independent solutions $\psi(x)$ and $\varphi(x)$. If $E$
is chosen to be real, then the wave functions can be chosen real
as well.
 The adjoint equation for $\varphi(x)$ is
\begin{equation}
- i \partial_x \varphi^{+}(x) \sigma_y  + \varphi^{+}(x) V(x) = E
\varphi^{+}(x)\,.
 \label{phi-adj}
\end{equation}
If we multiply \eref{dirac} by the row-vector $\varphi^+(x)$ from the left and
\eref{phi-adj} by the column-vector $\psi(x)$ from the right, the difference
takes the form
$ [W(\varphi, \psi)]_x = 0$
 (the subscript $x$
stands for the derivative with respect to $x$)
 meaning that the function
\begin{equation}
W(\varphi, \psi) =
\varphi^{+}(x) i \sigma_y \psi(x)\,. \label{dirac-wr}
\end{equation}
is (a non-zero) constant. Hence, $W(\varphi,\psi)$ here plays the
role of a Wronskian for the Dirac equation. Since $\psi$ and
$\varphi$ form a basis set for the two-dimensional solution space
of \eref{dirac} for any given $E$, any solution of \eref{dirac}
with the same $E$ may be expressed as
\begin{equation}
\chi(x) = A \psi(x) + B \varphi(x)\,, \label{com-sol}
\end{equation}
where $A,B\in\mathbb{C}$. Without loss of generality we may assume
that $\psi$ and $\varphi$ have been chosen so $W(\psi,\varphi)=1$.

Suppose now that $V(x)$ is a periodic potential:
\begin{equation}
V(x) = V(x+T)
 \quad T \in \mathbb{R}\,.
\label{Vperiodic}
\end{equation}
By Floquet's theorem, in this case \eref{dirac} has {\it{Bloch}}
solutions such that
\begin{equation}
\chi(x_0) =  \beta \chi(x_0+T) \quad \beta = \mbox{const} \in
\mathbb{C}\,, \label{qp-sol}
\end{equation}
where $x_0$ is any point on the real axis. By using
\eref{com-sol} we get the following equation for the coefficients
$A$ and $B$:
\begin{equation}
\psi(x_0) A + \varphi(x_0) B = \beta[\psi(x_0+T) A +
\varphi(x_0+T) B]\,. \label{ABeq}
\end{equation}
which is equivalent to the system of linear homogeneous equations:
\begin{equation}
\eqalign{
[\psi_1(x_0) - \beta \psi_1(x_0+T)] A +
[\varphi_1(x_0) - \beta \varphi_1(x_0+T) )]B = 0 \\
[\psi_2(x_0) - \beta \psi_2(x_0+T)] A + [\varphi_2(x_0) - \beta
\varphi_2(x_0+T) ]B = 0 }\,.
 \label{ABhomo}
\end{equation}
For such a system to have a non--trivial solution the determinant
of coefficients must vanish, which is the case only if $\beta$
satisfies the quadratic equation
\begin{equation}
\beta^2 - D \beta + 1 = 0\,, \label{beta-eq}
\end{equation}
where
\begin{equation}
D = W(\psi(x_0+T), \varphi(x_0)) + W(\psi(x_0), \varphi(x_0+T))
\label{lyap}
\end{equation}
and we have used the property $W(\psi ,\varphi )=1$. We get from
\eref{beta-eq} the two possible values for $\beta$
\begin{equation}
\beta_{1,2} = D/2 \pm \sqrt{D^2/4 - 1}\,,
 \quad \beta_1 \beta_2 = 1\,.
 \label{beta-12}
\end{equation}
This result is quite similar to that for the Schr\"odinger
equation with a periodic potential \cite{ejp_2001_v22_p22}. The
function $D=D(E)$ is called the {\it{Lyapunov function}} or {\it
Hill determinant} for the Dirac equation, and is real since $\psi$
and $\varphi$ are.

It is clear that for $|D(E)|<2$ the values of $\beta$ lie on the
unit circle of the complex $\beta$-plane, so we can write them
\begin{equation}
\beta_1 = \exp{(i K T)}\,, \quad \beta_2 = \exp{(-i K T)}\,.
\label{beta12com}
\end{equation}
with real $K$. It follows from (\ref{qp-sol}) that the Bloch
solutions are bounded on the whole real axis. Hence, all real  $E$
with $|D(E)|<2$ lie in allowed bands. Since for $|D(E)|>2$,
$\beta_{1,2}$ are real and positive, either Bloch solution is
unbounded. All such $E$ form the forbidden bands, while the band
edges are given by $|D(E)|=2$ and in general form the unbounded
sequence
\begin{equation}
\ldots E_{-4} < E_{-3} \leq E_{-2} < E_{-1} < E_1 < E_2 \leq E_3 <
E_4 \ldots \,.\label{band-edges}
\end{equation}
 An important feature
of the band structure for the Dirac equation is that allowed bands
may exist for negative energy.

\section{Darboux transformation operator}

The Darboux transformation provides a very powerful method for
finding new exactly solvable potentials, both for the
Schr\"odinger equation \cite{ppn-1997-v28-p374} and the Dirac
equation \cite{jpa-2002-v35-p3279}. In this section we shall
describe the main ideas for this method following
\cite{jpa-2002-v35-p3279,rjp-2000-v43-p48}.

Consider the stationary one-dimensional Dirac equation for the
Hamiltonian $h_0$:
\begin{equation}
 h_0 \psi(x) = (i \sigma_y \partial_x + V_0(x)) \psi(x) = E \psi(x)\,.
\label{dirac0}
\end{equation}
Suppose the solutions of  (\ref{dirac0}) are known for all $E$ and
we wish to solve a second Dirac equation having Hamiltonian $h_1$:
\begin{equation}
 h_1 \varphi(x) = (i \sigma_y \partial_x + V_1(x)) \varphi(x) = E \varphi(x)\,.
\label{dirac1}
\end{equation}
Instead of solving  (\ref{dirac1}) directly, one looks for a
{\it{transformation operator}} (or {\it{intertwiner}}) $L$ such
that
\begin{equation}
L h_0 = h_1 L\,. \label{intertw}
\end{equation}
If such an operator can be found, then the eigenspinors of $h_1$
may be obtained by applying $L$ to those of $h_0$:
$\varphi(x)=L\psi(x)$.

It has been shown (see \cite{jpa-2002-v35-p3279,rjp-2000-v43-p48}
for details) that a first order differential operator
\begin{equation}\label{L}
L=\partial_x-u'(x)u^{-1}(x)
\end{equation}
satisfies  (\ref{intertw}) with
\begin{equation}\label{V1}
V_1(x)=V_0+[i\sigma_y,u'(x)u^{-1}(x)]\,,
\end{equation}
where $u(x)=(u^{(1)}(x),u^{(2)}(x))$ is a $2\times 2$ matrix
satisfying the equation
\begin{equation}
h_0 u(x)  = u(x) \Lambda, \quad \Lambda = \mbox{diag}(\lambda_1,
\lambda_2)\,. \label{h0u}
\end{equation}
It is easy to check that the spinors $u^{(j)}(x)$, $j=1,2$ are
eigenfunctions of the Hamiltonian $h_0$ corresponding to the
eigenvalues $\lambda_j$. The matrix $u(x)$ is called a
{\it{transformation function}}.

If $E\ne \lambda_1$ or $\lambda_2$, then by applying $L$ to any
solution to  (\ref{dirac0}), we obtain a corresponding solution to
(\ref{dirac1}); if $E=\lambda_1$ or $\lambda_2$, then
$Lu^{(1,2)}(x)=0$. Nevertheless, in the latter case it can be
shown that the function
\begin{equation}
v(x)  = (u^t(x))^{-1} \label{v-u+}
\end{equation}
is a matrix eigenfunction for the Hamiltonian $h_1$ with
eigenvalue matrix $\Lambda$: $h_1v(x)=v(x)\Lambda.$ Therefore the
spinors $v^{(j)}(x)$, $j=1,2$ are eigenfunctions of $h_1$ with
eigenvalues $\lambda_j$. Solutions to (16), $\tilde{v}^{(1,2)}(x)$
for $E=\lambda_1$ and $E=\lambda_2$, may be found by means of the
properties
\begin{equation}
W(v^{(j)}(x),\tilde{v}^{(j)}(x)) = 1\,, \quad j=1,2 \,.
 \label{wd-v-vt}
\end{equation}

Therefore, with the help of the Darboux transformation one is able
to obtain the solutions to the transformed equation (\ref{dirac1})
for any value of $E$ provided the solutions to the original
equation (\ref{dirac0}) are known.

\section{Darboux transformation for a scalar potential}

A scalar potential is specified by a single function
$p_0(x)=m+S_0(x)$, $x\in \Bbb R$, so
\begin{equation}\label{scalar}
V_0 = (m +S_0(x)) \sigma_z = \left(\begin{array}{cc} m+S_0(x) &
0
\\ 0 & -(m+S_0(x))
\end{array}\right) \,,
\end{equation}
where $m$ is the particle mass. For our purposes, it is more
convenient to use the alternative form for the scalar potential
\begin{equation}\label{scalar1}
\tilde{V}_0 = (m +S_0(x)) \sigma_x = \left(\begin{array}{cc} 0 & m+S_0(x) \\
m+S_0(x) & 0
\end{array}\right) \,.
\end{equation}
These two potentials are related by the unitary transformation
\begin{equation}
\tilde{V}_0  = U^{-1} V_0 U\,, \tilde{\psi}  =  U \psi \, ,
\end{equation}
where
\begin{equation}
U = (1 + i \sigma_y)/ \sqrt{2}\, . \label{Umatr}
\end{equation}
In general, when a Darboux transformation is applied to a scalar
potential, the resulting potential is not scalar, but below we
formulate additional conditions that will prevent this (see
\cite{jpa-2002-v35-p3279,rjp-2002-v45-p74} for details).

If the spinor $u^{(1)}=(u_{11},u_{21})^t$ is a solution to the
Dirac equation (\ref{dirac0}) for the potential (\ref{scalar1}),
for some energy $E=\lambda$, it is easy to check that the spinor
$u^{(2)}=-\sigma_zu^{(1)}$ is a solution to the same equation for
energy $E=-\lambda$. The transformation matrix $u(x)$ constructed
from the spinors $u^{(1)}$ and $u^{(2)}$ can be seen to give
 \begin{equation}  \label{uxSC}
  u_x u^{-1}=
  \left(\begin{array}{cc}
  (\ln u_{11})' & 0\\
  0       & (\ln u_{21})'
  \end{array}\right)\,,
 \end{equation}
and the transformed potential (\ref{V1}) takes the form
 \begin{equation}
 \tilde{V}_1 = (m + S_1(x))\sigma_x\,,
 \label{V1sc}
 \end{equation}
 \begin{equation}
 S_1(x) =  S_0 + (\ln u_{21})' - (\ln u_{11})'
 \label{S1sc}
\end{equation}
which is clearly scalar.

With the help of  (\ref{L}) and (\ref{uxSC}) one finds the
solutions to the transformed equation:
  \begin{equation}
 \varphi = L \psi =  \left(\begin{array}{c}
  \psi'_1 - (\ln u_{11})' \psi_1\\
  \psi'_2 - (\ln u_{21})' \psi_2
 \end{array}\right).
 \label{sc_phi}
 \end{equation}
Note that the components of this spinor are given by the same
expressions that appear in the Darboux transformation, with
transformation functions $u_{11}$, $u_{21}$, for the Schr\"odinger
equation (see e.g. \cite{tmf-1995-v104-p356}).

\section{New periodic scalar potential}

The free particle Dirac Hamiltonian can be written in terms of the
scalar potential (\ref{scalar1}) with $S_0=0$. One can easily
verify that the spinors
\begin{equation}
\eqalign{
u^{(1)}(x) = \bigl( \cosh(\gamma x - \alpha)\,,
\cosh(\gamma x + \alpha)\bigr)^t,  \quad
e^{2\alpha} = \sqrt{\frac{m-\gamma}{m+\gamma}}~, \\
u^{(2)}(x) = - \sigma_z u^{(1)}(x)
}
\end{equation}
are eigenspinors of $h_0$ corresponding to the eigenvalues
$\lambda_{1,2}=\pm\lambda=\pm\sqrt{m^2-\gamma^2}$. By constructing
the transformation function from these spinors and using
(\ref{S1sc}), we arrive at the reflectionless (one-soliton)
potential corresponding to
\begin{equation}
S_1(x) = - \frac{2 \gamma^2}{m + \lambda \cosh 2 \gamma x}~,
\label{V1scalar}
\end{equation}
which was found previously  \cite{pra-1993-v47-p1708}. It is
easily checked that the two spinors given by the matrix
$(u^+)^{-1}$ are square integrable. Hence the potential
(\ref{V1sc}), (\ref{V1scalar}) has the two discrete levels
$E=\pm\lambda$.

Linearly independent solutions of the Dirac equation for potential
(\ref{V1scalar})  found by means of (\ref{sc_phi}) are
\begin{equation}  \label{t-psi}
 \tilde{\psi}(x) = \frac{E}{\sqrt{\gamma^2+k^2}}
\left( \begin{array}{l}
\cos kx - \frac{1}{k} w_1(x) \sin kx \\
\cos (kx-\delta) - \frac{1}{k} w_2(x) \sin (kx-\delta)
\end{array} \right) \, ,
\end{equation}
\begin{equation}
  \tilde{\varphi}(x) =
-\frac{1}{\sqrt{\gamma^2+k^2}} \left(
\begin{array}{l}
k \sin kx +  w_1(x) \cos kx \\
k \sin (kx-\delta) + w_2(x) \cos (kx-\delta)
\end{array} \right) \,,
\label{t-phi}
\end{equation}
where  $k = \sqrt{E^2 - m^2}$, $\delta = \arctan(k/m)$,
$w_{1,2}(x) = \gamma \tanh(\gamma x \mp \alpha)$. These are
normalized such that their Wronskian is unity:
$W(\tilde{\psi}_E(x),\tilde{\varphi}_E(x))=1$.

\begin{center}
\epsfig{file=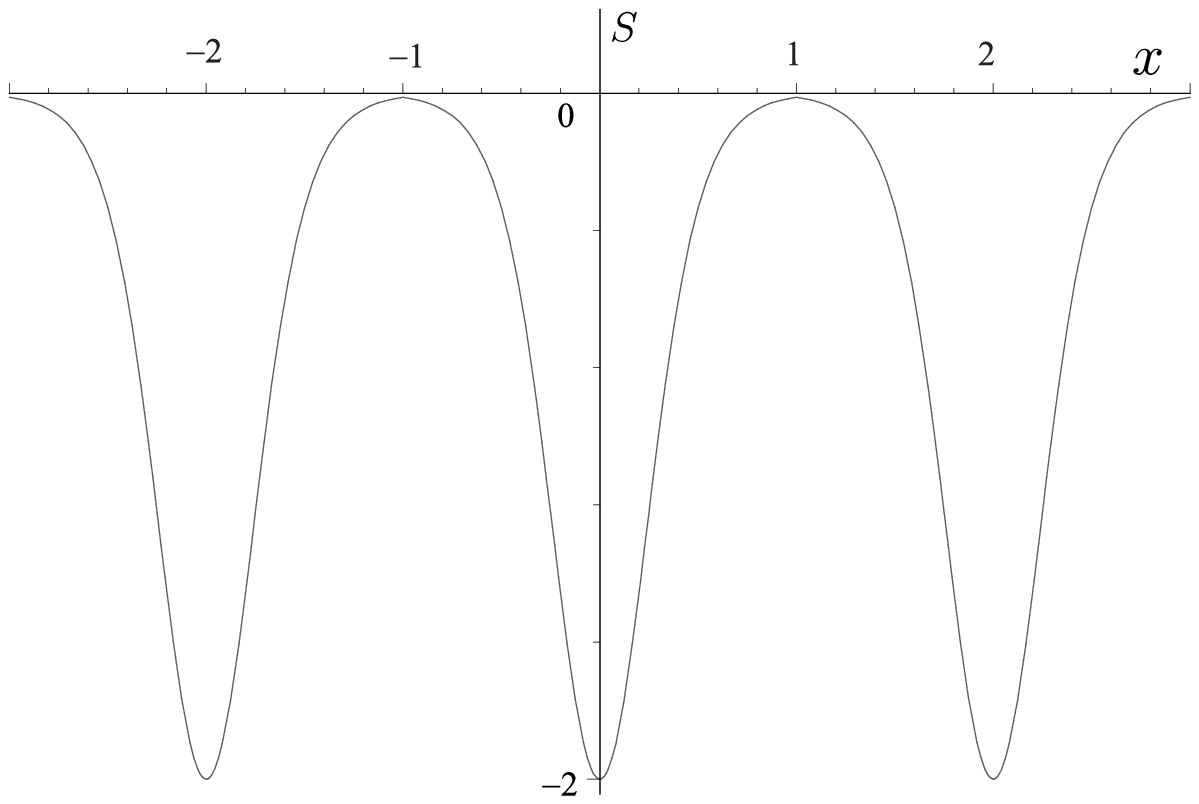, width=6cm}  
\label{fig1}
\medskip

\begin{minipage}{14cm}\begin{center}
{\small  Figure 1: Periodically continued scalar potential. }
\end{center}\end{minipage}
\end{center}
\medskip

Let us now restrict the potential as given by (\ref{V1scalar}) to
the interval $[-a,a]$, $a>0$ and continue it to the entire
$x$-axis by periodicity to obtain the continuous potential, having
period $T=2a$, shown in Fig.1 for $m=2$, $\lambda =1$, and $a=1$.
Having found the solution to the Dirac equation allows us to
calculate the Lyapunov function (\ref{lyap}). For this purpose we
first have to calculate Wronskians at $x_0=-a$. So, using
(\ref{t-psi}) and (\ref{t-phi}) after some simple algebra we find
for the first one the expression:
\begin{eqnarray}
\fl
W(\tilde \psi (a),\tilde \varphi (-a))=
 \frac{E}{
  k^2 +  \gamma^2    } \left[
    {w_1}(a)
    \left(\cos (2ka+\delta )\, -  \frac 1k
    \cos (ka+\delta )\,\sin (ka )\,{{w_1}(a)}\right)   \right.
\nonumber
\\
 \fl   \left.   +
k\,\cos (2ka)\,\sin (\delta ) - {w_2}(a)\,
   \left( \cos (2ka-\delta ) -
   \frac 1k
   \cos (ka)\,\sin (ka-\delta )\,{w_2}(a) \right)
   \right]
 \end{eqnarray}
 In the derivation we have used the symmetry relations
  $w_1(-a)=w_2(a)$,  $w_2(-a)=w_1(a)$.
After calculating the second Wronskian in  (\ref{lyap}) we obtain
\begin{eqnarray}
\fl D(E) = \frac{E}{k^2+\gamma^2} \bigl[ 2 w_1(a) \cos(2ka+\delta)
- 2 w_2(a) \cos(2ka-\delta) \nonumber \\
\lo+ \frac{1}{k}(k^2 - w_1^2(a))\sin(2ka+\delta) - \frac{1}{k}(k^2
- w_2^2(a))\sin(2ka-\delta) \bigr]\, . \label{V1lyapf}
\end{eqnarray}
The behavior of this function is sketched in Fig.2
\begin{center}
\epsfig{file=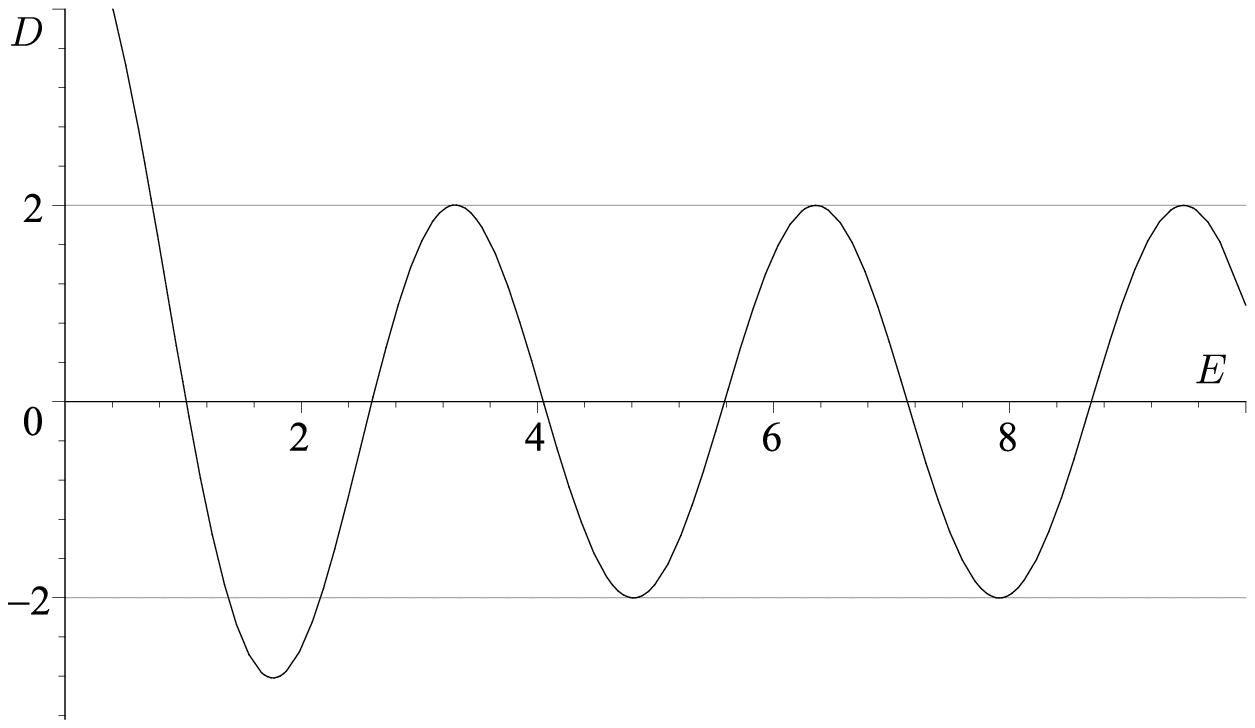, width=8cm}  
\label{fig1}
\medskip

\begin{minipage}{14cm}\begin{center}
{\small  Figure 2:  Lyapunov function for a periodic scalar
potential. }
\end{center}\end{minipage}
\end{center}
\medskip
 for $m=2$, $\lambda =1$, and $a=1$ and leads to the following
values for the lowest band edges in the positive spectrum: $E_0 =
0.738$, $E_1 = 1.381$, $E_2 = 2.164$, $E_3 = 3.274$, $E_4 =
3.335$, $E_5 = 4.802$, $E_6 = 4.827$, $E_7 = 6.352$. Since $D(E)$
is an even function of the energy, the negatives of these values
form the band edges in the negative spectrum.

We conclude this section by indicating how the actual energy bands
may be constructed from our results. From  \eref{qp-sol},
\eref{lyap} and \eref{beta12com}, since the Wronskian has the
value unity, we see that
 \begin{equation}
 \cos(2Ka)={\textstyle\frac{1}{2}}D(E)
 \end{equation}
  so that the wave vector $K$
is obtained as an explicit function of the energy in the $j$-th
energy band by
 \begin{equation}
 K={\textstyle \frac{1}{2a}}\arccos\; {\textstyle\frac{1}{2}}D(E) \mbox{
\hskip .2in for \hskip .1in} E_{j-1}<E<E_j\,.
 \end{equation}
 By symmetry
it is sufficient to consider only positive $K$, as we have
implicitly done here. In general the inversion of this to get $E$
as a function of $K$ (the so called dispersion law) must be
carried out numerically, but is trivially done graphically. The
lowest positive energy band for the model above is shown in Figure
3.
\begin{center}
\epsfig{file=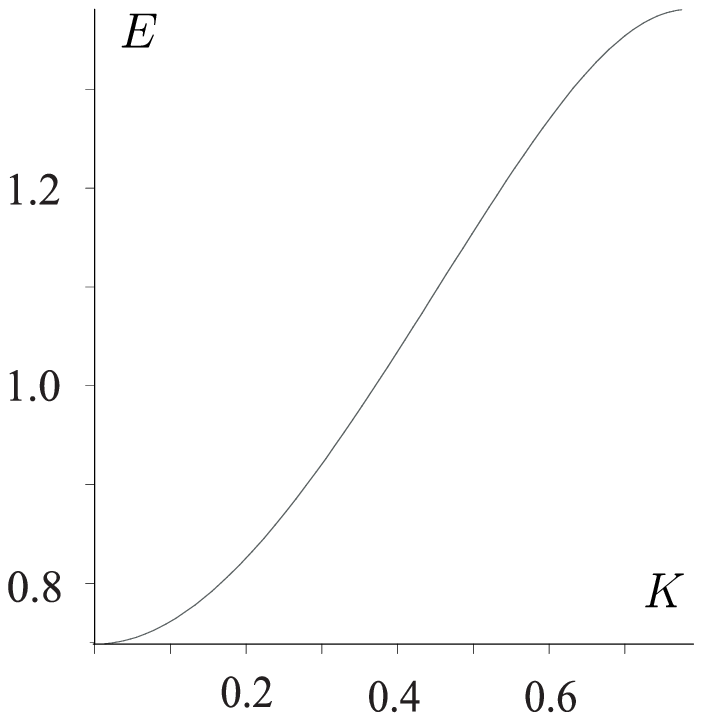, width=5cm}  
\label{fig1}
\medskip

\begin{minipage}{14cm}\begin{center}
{\small  Figure 3:  Lowest energy band for the case $\gamma=1$,
$m=2$
 for the scalar potential given by \protect\eref{V1scalar}. }
\end{center}\end{minipage}
\end{center}
\medskip

\section{Discussion}

In this note, using a specific example of one-soliton potential,
we have shown how a great variety of exactly solvable continuous
periodic local potentials for the one dimensional Dirac equation
may be
 constructed. Furthermore,
this procedure may be manipulated so that only elementary
mathematical functions are involved in the construction and in the
solution. Up till now, the only examples available have been
simple variants of the classic Kronig-Penney model, yet this has
been a key test bed in the relativistic theory of surface states
\cite{12}, for example. We feel that the class of models presented
here will be equally useful and lead to new insights in this and
other areas.

\ack
The work of BFS was partially supported by the Spanish MCYT and
the European FEDER (grant BFM2002-03773), and also by Ministerio
de Educaci\'on, Cultura y Deporte of Spain (grant SAB2000-0240).
MLG thanks the Universidad de Valladolid for hospitatlity and
support. He also acknowledges partial support from the NSF (USA)
under grant DMR-0121146.


\end{document}